\documentclass[11pt]{article}
\usepackage{latexsym}  
\usepackage{amssymb}
\usepackage{graphicx}
\usepackage{amsmath}

\newcommand{\be}{\begin{equation}}
\newcommand{\ee}{\end{equation}}
\newcommand{\bea}{\begin{eqnarray}}
\newcommand{\eea}{\end{eqnarray}}
\newcommand{\bref}[1]{(\ref{#1})}
\newcommand{\la}{\langle}
\newcommand{\ra}{\rangle}
\begin{document}
\begin{titlepage}
\begin{flushright}
\today
\end{flushright}
\vspace{4\baselineskip}
\begin{center}
{\Large\bf Filling the gaps between model predictions and their prerequisites in electric dipole moments.}
\end{center}
\vspace{1cm}
\begin{center}
{\large Takeshi Fukuyama$^{a,}$
\footnote{E-mail:fukuyama@se.ritsumei.ac.jp}}
and
{\large Koichiro Asahi$^{b,}$
\footnote{E-mail:asahi@phys.titech.ac.jp}}
\end{center}
\vspace{0.2cm}
\begin{center}

${}^{a} $ {\small \it Research Center for Nuclear Physics (RCNP),
Osaka University, Ibaraki, Osaka, 567-0047, Japan}\\[.2cm]

${}^{b}$ {\small \it Department of Physics, Tokyo Institute of Technology, 2-12-1 Oh-okayama, Meguro, Tokyo 152-8551, Japan}

\vskip 10mm
\end{center}
\vskip 10mm
\begin{abstract}
We clarify the conditions or assumptions under which theoretical predictions of various models beyond the standard model give mainly in electric dipole moments. The correct interpretation of those conditions seems to be indispensable to the refinements of model building as well as to the mutual reliance in experimental and theoritical communities. The connections of these analyses to the recent experimental results at the LHC and the other places are also discussed. 
\end{abstract}
\end{titlepage}
\section{Introduction}
Parmanent electric dipole moments (EDMs) of particles are smoking guns of new physics beyond the standard model (BSM). Indeed, the standard model (SM) predicts EDMs of all particles far beyond the upper bounds of ongoing and near-future experiments, whereas many models beyond the SM (BSM models) suggest many orders of magnitudes larger than those of the SM. Some of them seem to give even larger values than the existing upper bound of experiments. Another peculiar property of EDM is that it appears in a variety of hierarchical stages of matter. For instance, electron EDM appears also in paramagnetic atoms and molecules in enhanced forms. This drives us to study, beyond particle physics, a wide range of fields, nuclear physics, atomic physics, chemical physics, and solid state physics.
These properties, a variety of models and a variety of EDM appearances, motivated us to write the review of EDM which resumes these diverse fields in a compact and self-complete form \cite{Fuku1}. However, experimental developments are beyond our expectation. Especially, the recent improvement of the upper bound on electron EDM by ACME \cite{ACME} and the discovery of $126$ GeV Higgs particle and the negative searches for SUSY particles by the Large Hadronic Collider (LHC) at CERN are especially impressive \cite{ATLAS} \cite{CMS}. They give rich precious informations to models and force them to be modified.  There have arisen many discussions between the experimental and theoretial communities. In these situations it is very useful if many scientists over the wide regions can easily see the list of predictions made by various models. One of the most excellent and well known one may be Figure 1 by Pendlebury and Hinds \cite{Hinds}. 
\begin{figure}[h]
\begin{center}
\includegraphics[scale=0.2]{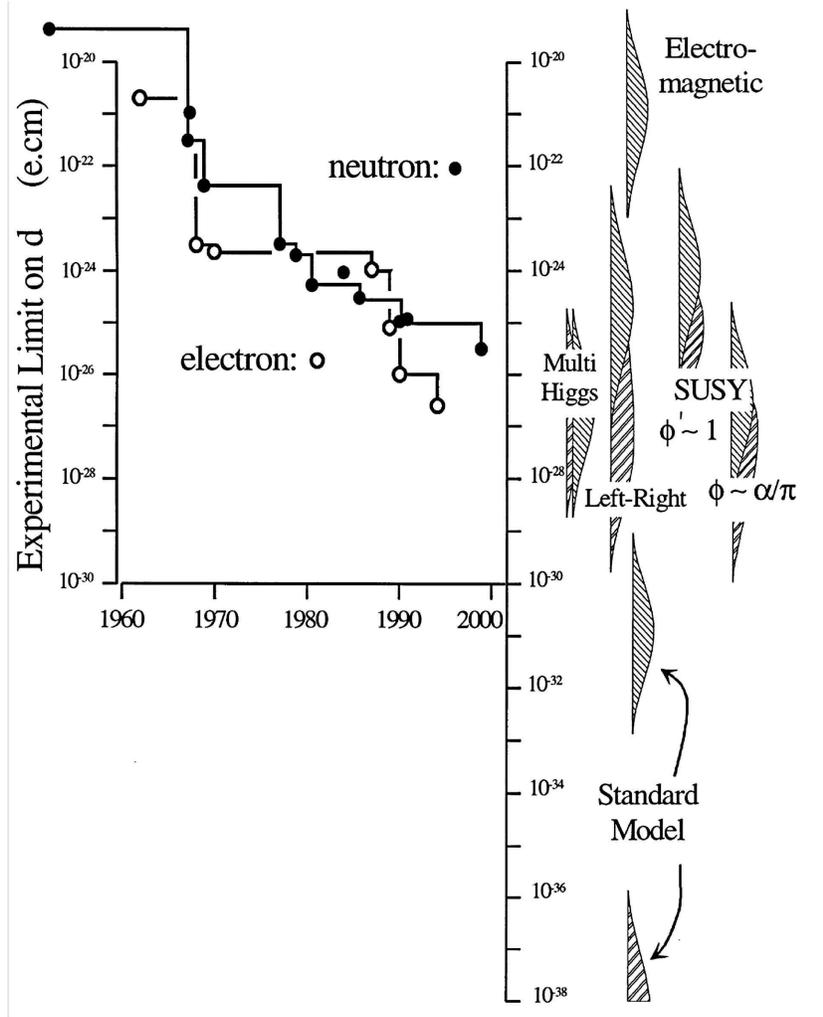}
\end{center}  
  \caption{Graph on the left:History of experimental upper limits on the electron and neutron EDM. Right: Characterestic EDM predictions in various particle theory models. Predictions for the electron EDM tend to lie lower than for the neutron and are hatched with double lines cited from \cite{Hinds} by permision.}
\end{figure}
Unfortunately, we have seen very frequently that so many peoples cite these model results without carefully considering the conditions or assumptions under which their results are obtained.  These situations are not happy for the mutual understanding between experimental and theoretical communities, and model building itself. This is the motivation of this letter. The main objects of this letter, therefore, are to clarify the above mentioned conditions, referring the details of calculations to \cite{Fuku1}, and to add some comments on the connections to the recent great experimental results.

\section{EDMs in the SM}
In the SM, EDMs come from Kobayashi-Maskawa CP violating phase \cite{KM}, apart from possible very small contribution from $\theta$ term in quark sector.
We resume in Table I the predicted EDM values of various particles in the framework of the SM and experimental results. The detailed explanations of it are given in \cite{Fuku1}.

\begin{table}[h]
\caption{The resume of EDM values or bounds from theoretical and experimental sides.
The second, third, and fourth columns are the number of loops in the first graph giving nonzero value in the SM, EDM value in the SM, and experimental upper bounds, repectively. (*) The SM bound of nucleons from Schiff moment of Hg is much weaker than those of 3-loop diagram, $d_n< 4.0\times 10^{-25},~d_p<3.8\times 10^{-24}$ e cm \cite{Dmitriev}.}
{\begin{tabular}{|c|c|c|c|}
\hline
particle &loop number& EDM value in SM/(e cm)& experimental upper bound/(e cm)\\
\hline
quark&$3$&$d_d\approx 10^{-34}$& \\
\hline
neutron&$3$&$d_n\approx 10^{-32}$ \cite{Seng}(*)& $2.9\times 10^{-26}$ \cite{Baker} \\
\hline
proton&$3$&$d_p\approx 10^{-32}$ (*) & $7.9\times 10^{-25}$ \cite{Hg} \\
\hline
deuteron& & $d_D\approx 1.5\times 10^{-31}$ \cite{Yamanaka}&  \\
\hline
W boson&$3$&$d_W\approx 8\times 10^{-30}$\\
\hline
lepton&$4$& $|d_e|\approx 8\times 10^{-41}$& $|d_e|\leq 8.7\times 10^{-29}$ \cite{ACME}\\
& & &$|d_\mu|=1.9\times 10^{-19}$\cite{Beringer}\\
\hline
\end{tabular}}
\end{table}
As a result of loop corrections, the effective Lagrangian is written as
\begin{eqnarray}
&&
 -i\overline{\psi}_i
 \left(
  A_L^{ij} P_L + A_R^{ij} P_R
 \right)
 \sigma^{\mu\nu}
 \psi_j
 F_{\mu\nu}\nonumber\\
&&\hspace*{5mm}
=
 \frac{-i}{\,2\,}
 (A_L^{ij} + A_R^{ij})
 \overline{\psi}
 \sigma^{\mu\nu}
 \psi
 F_{\mu\nu}
 +
 \frac{1}{\,2\,}
 (A_R^{ij} - A_L^{ij})
 \overline{\psi}
 \sigma^{\mu\nu} \gamma^5
 \psi
 F_{\mu\nu}\nonumber\\
&&\hspace*{5mm}
=
 (A_L^{ij} + A_R^{ij})
 \overline{\psi}
 \left(
  \begin{array}{cc}
   \vec{\sigma}\cdot\vec{B} & 0\\
   0 & \vec{\sigma}\cdot\vec{B}
  \end{array}
 \right)
 \psi
 +i
 (A_R^{ij} - A_L^{ij})
 \overline{\psi}
 \left(
  \begin{array}{cc}
   0 & \vec{\sigma}\cdot\vec{E}\\
    \vec{\sigma}\cdot\vec{E} & 0
  \end{array}
 \right)
 \psi,
\label{eff_dipole}
\end{eqnarray}
where notations follow those in \cite{Fuku1}. 
 For the EDM and magnetic dipole moment (MDM),
we take zero momentum of the photon.
 Then imaginary part of coefficients
of the effective interaction vanishes
because of the optical theorem
(imaginary part of the forward scattering amplitude
is given by the sum of possible cuts of intermediate states).
 We have an anomalous MDM $a_\psi$
and EDM $d_\psi$ of particle $\psi$ as
\begin{eqnarray}
a_\psi
&=&
 \frac{g-2}{2}
=
 - \frac{2 m_\psi}{ \text{e} Q_\psi }\,
   \Re ( A_R^{ii} + A_L^{ii} ),\\
d_\psi
&=&
 2\,\Im ( A_R^{ii} - A_L^{ii} ).
\label{imaginary}
\end{eqnarray}
Here $\Re$ and $\Im$ express taking the real and imaginary parts, respectively. 

For neutron EDM, we must take a contribution of chromo EDM $\tilde{d}$ \cite{Pospelov2}
\be
d_n=(1\pm 0.5)\frac{|<\overline{q}q>|}{225\mbox{MeV}^3}\left[0.55e(\tilde{d}_d+0.5\tilde{d}_u)+0.7(d_d-0.25d_u)\right].
\label{d_n}
\ee
The experimental bound on neucleon may be obtained from the experimental value of ${}^{199}$Hg EDM \cite{Hg},
\be
d_{\mbox{Hg}}\leq 3.1\times 10^{-29} ~\mbox{[e cm]}~~(95\%~C.L.)
\ee
and theoretical estimations
\be
d_{\mbox{Hg}}=-(\tilde{d}_d-\tilde{d}_u-0.012\tilde{d}_s)\times 3.2\times 10^{-2}e
\ee
by the QCD sum rule \cite{Falk} and
\be
d_{\mbox{Hg}}=-(\tilde{d}_d-\tilde{d}_u-0.0051\tilde{d}_s)\times 8.7\times 10^{-3}e
\ee
by the chiral Lagrangian method \cite{Hisano}.

EDMs of light nuclear systems are calculated by combining effective $\Delta S=1$ four quark interactions and $\pi,~\eta$ interaction \cite{Yamanaka}.

For electron EDM there are many ongoing and near-future ambitious experiments using polarized molecules. The advantage of diatomic molecule is due to strong internal electric field and close rotation-vibration energy levels etc. \cite{Fuku1, Hinds2, DeMille}. So experimental $d_e$ will severely constrain BSM models in near-future. 

\section{EDMs in New Physics}
In this section, we explain on the model predictions and thir prerequisites in the typical models, two Higgs doublet model, left-right symmetric models, MSSM, and SUSY SO(10) GUT models. 
\subsection{Two Higgs Doublet Models (2HDMs)}
As the simplest extension of the Higgs sector of the SM
which has only one Higgs doublet $\phi_1$,
another Higgs doublet $\phi_2$ is introduced
in the 2HDMs \cite{2HDM, Djouadi, Branco}. \footnote{We may consider the triplet Higgs model as a simpler model than 2HDM. However, it does not give any predictable CP phase and we do not discuss here.}
 There are several types of the model
depending on which doublet couples with which fermion:
\begin{eqnarray}
\text{type I (SM-like)}
&:&
 \text{$\phi_1$ couples with all fermions}\nonumber\\
&&
 \text{$\phi_2$ decouples from fermions}\nonumber\\
\text{type II (MSSM-like)}
&:&
 \text{$\phi_1$ couples with down-type quarks and charged leptons}\nonumber\\
&&
 \text{$\phi_2$ couples with up-type quarks}\nonumber\\
\text{type III (general)}
&:&
 \text{both of Higgs doublets couple with all fermions}\nonumber\\
etc. && \nonumber
\end{eqnarray}
$\phi_i$ have vacuum expectation values (vevs) $\la \phi_i\ra=v_i~i=1,2$ and they are rotated by $\beta=\tan^{-1}v_2/v_1$ as
\be
\Phi_1=\left(
\begin{array}{cc}
G^+\\
\frac{v+H_1+iG^0}{\sqrt{2}}\\
\end{array}
\right),
~~~
\Phi_2=\left(
\begin{array}{cc}
H^+\\
\frac{H_2+iA}{\sqrt{2}}\\
\end{array}
\right)
\ee
with $v=\sqrt{v_1^2+v_2^2}$. Here $H_i$ and $A$ are CP-even and CP-odd, respectively.
Fig. 1 contributes to EDM via 
\bea
\la H_1A\ra&=&\frac{1}{2}\sum_n\frac{\sin 2\beta \Im Z_{0n}}{q^2-m_{n}^2},\nonumber\\
\la H_2A\ra&=&\frac{1}{2}\sum_n\frac{\cos 2\beta \Im Z_{0n}-\Im\tilde{Z}_{0n}}{q^2-m_{n}^2},
\eea
where the summation is over all the mass eigenstates of neutral Higgs bosons, and the explicit forms of $Z_{0n}$ and $\tilde{Z}_{0n}$ are given in \cite{Weinberg}.

One-loop diagram (Fig. 2 (a)) contribution irrelevant to strong interaction is 
\bea
d_{e,\mu}^{1-{\rm loop}}
= \frac{ e \sqrt{2} G_F \tan^2\beta }{ (4\pi)^2 }
      (m^3/m_0^2)[\mbox{ln}(m^2/m_0^2)+3/2]( \Im Z_0 +\Im\tilde{Z}_0 ) 
\eea
in the limit of $m^2/m_0^2\ll 1$. Here $m_0$ is the mass of the lightest neutral Higgs boson and $m=m_e,~m_\mu$ \cite{Barger}.
\begin{figure}[h]
\begin{center}
\includegraphics[scale=0.8]{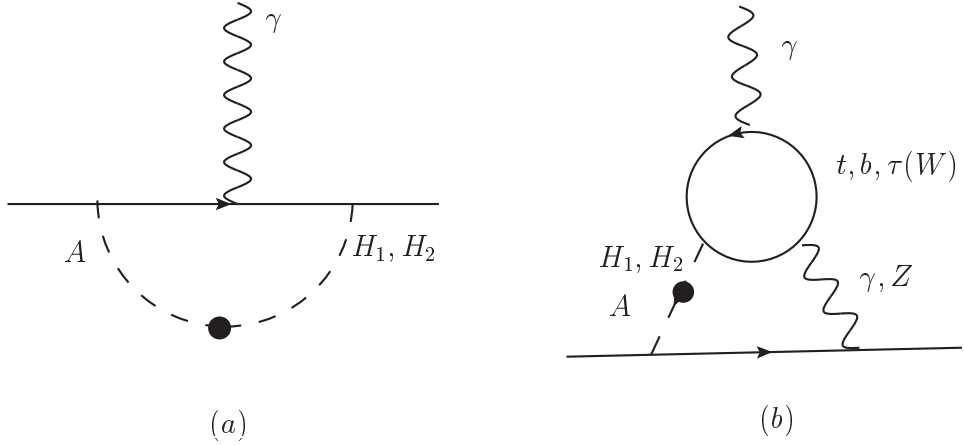}
\end{center}  
  \caption{Diagrams contributing to EDMs of electron and muon in 2HDM.
}\label{fig3}
\end{figure}

However two-loop Barr-Zee diagram (Fig. 2 (b)) dominates over the one-loop diagram \cite{Barr}. Dominant top loop diagram, for instance, gives
\be
d_{e,\mu}^{\mbox{top}}=-\frac{16}{3}\frac{em_{e,\mu}\alpha\sqrt{2}G_F}{(4\pi)^3}\{[f(r_t)+g(r_t)]\Im Z_0+[g(r_t)-f(r_t)]\Im\tilde{Z}_0\}
\label{demu}
\ee
with $r_t=m_t^2/m_0^2$. Here loop functions are given by
\bea
f(r)&=&\frac{r}{2}\int_0^1dx\frac{1-2x(1-x)}{x(1-x)-r}\mbox{ln}\left[\frac{x(1-x)}{r}\right], \\
g(r)&=&\frac{r}{2}\int_0^1dx\frac{1}{x(1-x)-r}\mbox{ln}\left[\frac{x(1-x)}{r}\right]
\eea
and
\be
\frac{em\alpha\sqrt{2}G_F}{(4\pi)^3}\approx 1\times 10^{-25}(m/(0.1\mbox{GeV})) ~\mbox{[e cm]}.
\label{2HDMvalue}
\ee
The other loop contributions are given in \cite{Barger} \cite{Jung}.\\
{$\clubsuit$ \bf Condition;}\\
By comparing Eq.\bref{demu} with Fig.1, you can easily see that the estimated values in \bref{2HDMvalue} and in \cite{Hinds} are in units of $\Im Z_0$. However, it is probable that $|\Im Z_0|\ll 1$. Indeed, the masses of neutral and charged Higgses and phases are tightly constrained from
 $R_b\equiv \frac{\Gamma(Z\rightarrow b\overline{b})}{\Gamma(Z\rightarrow hadrons)},~
\Gamma(b\rightarrow s\gamma),~\overline{B}^0-B$ mixing, $\rho$ parameter etc., and we should take those constraints into account. Also $Z_0$ satisfies the sum rule \cite{Weinberg}
\be
\sum_n \tilde{Z}_{n0}=\sum_n Z_{n0}=0.
\ee
Even if we assumed  $|\Im Z_0|=1$, the estimated value amounts to be O($10^{-27}$) [e cm] for electron EDM. However, it is true that the recent experiment \cite{ACME} gives severe constraints on the type II model.
It should be remarked that EDM is linear in its mass for two-loop and cubic for one-loop diagrams. 
Here we have given the formulae for the type II model. It can be easily modified for the type I and the other models \cite{Abe} in which the numerical values of chromo EDMs in 2HDM are also given under special assumptions of four unknown Higgs self coupling constants.
One of the most typical processes to check 2HDM is the $4\sigma$ excess of tauonic B decay from the SM, $
\overline{B}\rightarrow D^{(*)}\tau\overline{\nu}_\tau$.
Unfortunately, however, the type II model (and probably the other types also) may be excluded at least in their naive forms from the experiment, that is, mismatch of $R(D)$ and $R(D^*)$ in terms of $\tan\beta/m_{H^+}$ \cite{Babar}. Here
\bea
R(D)\equiv \frac{Br(\overline{B}\rightarrow D\tau^-\overline{\nu}_\tau)}{Br(\overline{B}\rightarrow D\l^-\overline{\nu}_l)}
\eea
and $l$ is either $e$ or $\mu$. Also we need more constraints by applying the model to many phenomena. In this sence, recent indications of diphoton excess at $750$ GeV found at the LHC \cite{diphoton1, diphoton2} might be a good chance for it.
This is because 2HDM has additional neutral Higgses, $H_0,~A_0\equiv \chi$. Unfortunately these contributions to $\chi\to \gamma\gamma$ come from top quark loop and severly suppressed by $m_t\ll m_\chi\approx 750$ GeV \cite{Angelescu}. In order to realise the observed data, in most cases, heavy vectorlike fermions are introduced. This is the very drastic change of physics. We need more definite and broad events supporting such fermions.
In this sence more precise values of resonance at $750$ GeV.

\subsection{Left-Right (LR) Symmetric Models}
There are many LR models.
The smallest gauge group of LR symmetry is
\bea
SU(2)_L\times SU(2)_R\times U(1)_{B-L}
\eea
Then charge quantization \cite{PS}\cite{MP},
\bea
Q=I_{3L}+I_{3R}+\frac{1}{2}(B-L),
\label{chargequant}
\eea
is realized.
If we consider it as a remnant from SO(10), $SO(10)\rightarrow SU(4)_c\times SU(2)_L\times SU(2)_R$ (If we start from this group, it is called the Pati-Salam (PS) model.), it satisfies at a certain energy scale $v_{\mbox{PS}}$
\bea
g_L=g_R
\eea
and PS model is unified at $M_{GUT}$ as
\bea
\frac{M_4}{\alpha_4}=\frac{M_{2L}}{\alpha_{2L}}=\frac{M_{2LR}}{\alpha_{2R}}=\frac{M_{1/2}}{\alpha_{GUT}}.
\label{PS}
\eea
Also mixing matrices of left-handed and right-handed fermions are the same.
Of course, these constraints are realized at $v_{\mbox{PS}}$ but break down as the energy goes down to the SM scale by renormalization effects.
However, we do not assume such scheme here.  Matters and Higgs are assinged as
\bea
\Phi
\equiv
\left(
 \begin{array}{cc}
  \phi_1^0 & \phi_2^+\\
  \phi_1^- & \phi_2^0
 \end{array}
\right)=(1,2,2,0),
\eea
\bea
Q_L=\left(
\begin{array}{cc}
u_L^i \\
d_L^i
\end{array}
\right) =(3,2,1,1/3),~~~
Q_R=\left(
\begin{array}{cc}
u_R^i \\
d_R^i
\end{array}
\right) =(3,1,2,1/3)
\eea
under $SU(3)_c\times SU(2)_L\times SU(2)_R\times U(1)_{B-L}$. 
Left and right-handed doublets of lepton, $L_L$ and $L_R$, having the quantum number $(1,2,1,-1)$ and $(1,1,2,-1)$, are also incorporated.

$\Phi$ couples with $\overline{Q}_LQ_R$, so $(B-L)(\Phi)=0.$
This and \bref{chargequant} indicate that the symmetry breaking 
\bea
\langle \Phi\rangle=\left(
\begin{array}{cc}
\kappa & 0 \\
0 & \kappa'e^{i\lambda}
\end{array}
\right)
\eea
leads to $U(1)_{I_{3L}+I_{3R}}\times U(1)_{B-L}$
and not to $U(1)_Q$ \cite{Pal}. So we need additionally, for instance, 
\bea
\Delta_L=(3,1,2), ~~\Delta_R=(1,3,2).
\eea
Then the mass matrix of charged L-R weak bosons becomes
\bea
\left(
\begin{array}{cc}
\frac{1}{2}g^2(\kappa^2+\kappa'^2+2v_L^2) & g^2\kappa\kappa'\\
g^2\kappa\kappa' & \frac{1}{2}g^2(\kappa^2+\kappa'^2+2v_R^2)
\end{array}
\right),
\eea
where $v_L$ and $v_R$ are vevs of $\Delta_L$ and $\Delta_R$, respectively ($v_R\gg v_L$). The transformation angle $\zeta$ from $W_{L,R}$ to mass eigenstates $W_{1,2}$,
\be
W_1=W_L\cos \zeta -W_R\sin \zeta e^{i\lambda},~~W_2=W_L\sin\zeta e^{-i\lambda}+W_R\cos\zeta, 
\label{LR3}
\ee
is given by
\bea
\mbox{tan}2\zeta=\frac{2\kappa\kappa'}{v_R^2-v_L^2}\approx \frac{M_{W_L}^2}{M_{W_R}^2}.
\eea
\begin{figure}[h]
\begin{center}
\includegraphics[scale=0.8]{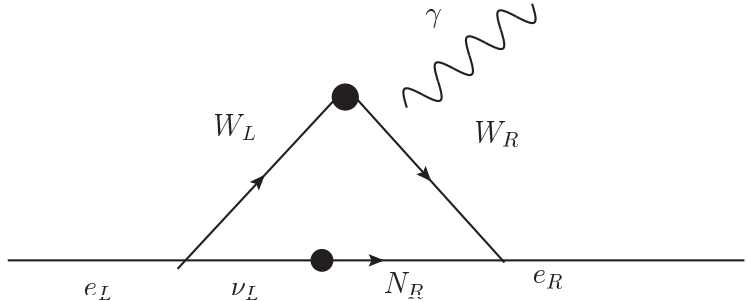}
\end{center}  
  \caption{LR symmetric diagram giving rise to electron EDM. 
}\label{LR}
\end{figure}

Lepton EDM comes mainly from the diagram (Fig. 3) and is given by \cite{Suzuki}
\bea
&&d_e=\frac{eG_F}{8\sqrt{2}\pi^2}I_1(\frac{m_R^2}{M_{W_L}^2},0)\mbox{sin}2\zeta\Im (m_D)\\
&&=2.1\times 10^{-24}I_1(\frac{m_R^2}{M_{W_L}^2},0)\mbox{sin}2\zeta (\Im (m_D)/1 \mbox{MeV})~\mbox{[e cm]}.
\eea
Here $m_R$ ($m_D$) is the mass of heavy right-handed neutrino $N_R$ (Dirac neutrino), and the mass of light left-handed neutrino $\nu_L$ is given by $m_{\nu}\approx m_D^2/m_R$ for the type I seesaw. The loop function $I_1$ is
\be
I_1(r,0)\approx \frac{2}{(1-r)^2}\left(1-\frac{11}{4}r+\frac{1}{4}r^2-\frac{3r^2\mbox{ln}r}{2(1-r)}\right).
\ee 
Inserting the observed upper limit of
\bea
|\mbox{sin}2\zeta |<2.74\times 10^{-3}~\mbox{or}~\zeta\approx\frac{M_{W_L}^2}{M_{W_R}^2}
\label{zeta}
\eea
with
\bea
M_{W_R}>2.15\times 10^3~~\mbox{GeV 95\%C.L.},
\eea
we obtain
\be
|d_e|
<  2.8\times 10^{-27}\,
   \frac{|\Im(m_D)|}{\rm{MeV}}\,~ \rm{e~ cm} \quad
   \rm{for} \left( \frac{m_R}{M_{W_L}} \right)^2 \gg 1.
\label{LRvalue1}
\ee
Here we have considered a TeV order seesaw. It is more probable that $m_D$ is of order of top quark masss but in this case $v_R$ amounts to O($10^{13}$) GeV and $d_e$ is much less than the value of \bref{LRvalue1}.
The LR model also predicts $d_n$ \cite{Mohapatra} utilizing the CP parameters of the neutral kaon system,
\be
d_n\approx (10^{-21}\mbox{[e cm]})|\eta_{+-}-\eta_{00}|\frac{4\sin \lambda+1.4\sin (\lambda-\delta)-0.1\sin (\lambda+\delta)}{\left(\sin\lambda+\sin (\lambda+\delta )\right)|\gamma|}
\label{LRvalue2}
\ee
via
\be
|\eta_{+-}-\eta_{00}|\approx \tan \zeta(\sin\lambda+\sin (\lambda+\delta))|\gamma|.
\label{zeta2}
\ee
Here quark charge current eigenstates $(d^0,~s^0)$ are transformed to mass eigenstatres $(d,~s)$, generalizing to invoke complex vacuum expectation value,
\be
d_{L,R}=d^0_{L,R}\cos\theta_{L,R}-s^0_{L,R}\sin\theta_{L,R}e^{i\delta_{L,R}},~~s_{L,R}=d^0\sin\theta_{L,R}e^{-i\delta_{L,R}}+s^0\cos\theta_{L,R} 
\label{LR2}
\ee
and $\delta=\delta_L-\delta_R$. $\gamma$ is an O(1) numerical factor coming from strong interaction.\\
$\clubsuit$ {\bf Condition;}\\
The expression \bref{LRvalue2} has simple forms in the following two cases:\\
(i)  $\delta=0$
\be
|d_n|\approx (2.7/||\gamma|)\times (10^{-21}\mbox{e cm})|\eta_{+-}-\eta_{00}|
\ee
and \\
(ii)  $\lambda=0$
\be
|d_n|\approx (1.5/||\gamma|)\times (10^{-21}\mbox{e cm})|\eta_{+-}-\eta_{00}|
\ee
Due to PDG2012 \cite{Beringer}
\be
|\eta_{00}|=(2.222\pm0.010)\times 10^{-3},~~|\eta_{+-}|=(2.233\pm0.010)\times 10^{-3}.
\label{eta}
\ee
Using \bref{zeta2} and \bref{eta}, we obtain
\be
|\zeta|\geq 10^{-5}/|\gamma|\geq 10^{-6}
\ee
This is consistent with \bref{zeta}. As for the recent diphoton excess at $750$ GeV, the situation is same as the case of 2HDM.
\subsection{Minimal Supersymmetric Standard Model (MSSM)}
MSSM has many CP phases and EDM appears in one-loop level. Note that $A_L$ and $A_R$ in \bref{eff_dipole} must include a fermion mass (or sfermion mass in the loop)
because the effective interaction
$\overline{\psi} \sigma^{\mu\nu} \psi$
changes the chirality which can be done
by the mass term in the soft SUSY breaking Lagrangian \cite{Martin},
\bea
\label{ss}
{\cal L}_{soft}&=&-\frac{1}{2}\left(M_3\tilde{g}\tilde{g}+M_2\tilde{W}\tilde{W}+M_1\tilde{B}\tilde{B}+c.c.\right)\nonumber\\
&-&\left(\tilde{\overline{u}}{\bf A}_u\tilde{Q}H_u-\tilde{\overline{d}}{\bf A}_d\tilde{Q}H_d-\tilde{\overline{e}}{\bf A}_e\tilde{L}H_d+c.c.\right)\\
&-&\tilde{Q}^\dagger {\bf m}_Q^2\tilde{Q}-\tilde{L}^\dagger {\bf m}_L^2\tilde{L}-\tilde{\overline{u}}^\dagger {\bf m}_u^2\tilde{\overline{u}}^\dagger-\tilde{\overline{d}}{\bf m}_d^2\tilde{\overline{d}}^\dagger-\tilde{\overline{e}}{\bf m}_e^2\tilde{\overline{e}}^\dagger\nonumber\\
&-&m_{H_u}^2H_u^*H_u-m_{H_d}^2H_d^*H_d-\left(bH_uH_d+c.c.\right).\nonumber
\eea
Here tilde marks the SUSY partner. So even if we do not incorporate new CP phase, we have 19 parameters (3 gaugino masses + tan$\beta$ + $\mu$ + $b$ + 10 sfermion masses + 3 trilinear terms) in addition to the SM+neutrino parameters ($27$ for Dirac $\nu$  or $29$ for Majorana $\nu$), called phenomenological MSSM (pMSSM). 
Universal SUSY-breaking is a very strong assumpotion that it requires not only flavour-blindness but also universality over quarks and leptons at GUT scale,
\bea
&&{\bf m}_Q^2={\bf m}_{\overline{u}}^2={\bf m}_{\overline{d}}^2={\bf m}_{\overline{L}}^2={\bf m}_{\overline{e}}^2=m_0^2{\bf 1}_3,\label{USB1}\\
&&m_{H_u}=m_{H_d}=m_0,\label{NUHM}\\
&&\frac{M_3}{g_3^2}=\frac{M_2}{g_2^2}=\frac{M_1}{g_1^2}=\frac{M_{1/2}}{g_u^2},\label{USB2}\\
&&{\bf A}_u=A_0{\bf Y}_u,~~{\bf A}_d=A_0{\bf Y}_d,~~{\bf A}_e=A_0{\bf Y}_e.\label{USB3}
\eea
with five parameters, $m_0,~M_{1/2},~A_0,~\tan\beta,~\mbox{sign}(\mu)$.
This MSSM+universal SUSY breaking boundary condition is called constrained MSSM (CMSSM).
If, in place of \bref{NUHM}, we set $m_{H_u},~m_{H_d}$ as free parameters, it is called non-universal Higgs masses (NUHM2) model \cite{E-O-S} or called NUHM1 for $m_{H_u}=\pm m_{H_d}$ \cite{Baer}.
Lepton ($l$) EDMs ($d_l$) come from one-loop diagram (Fig. 4) and are \cite{Kingman}
\be
\left(\frac{d_l}{e}\right)^\chi=\frac{M_{\chi_i}}{16\pi^2m_{\tilde{l}'_j}}\Im \left[\left(g_{Rij}^{\chi l\tilde{l}'}\right)^*g_{Lij}^{\chi l\tilde{l}'}\right]\left[Q_\chi A(M_{\chi_i}^2/m_{\tilde{l}'_j}^2)+Q_{\tilde{l}'}B(M_{\chi_i}^2/m_{\tilde{l}'_j}^2)\right],
\ee
where $A$ and $B$ are loop fuctions of chargino and neutralino contributing diagrams, respectively, whose explicit forms with those of coupling of $\left(g_{Rij}^{\chi l\tilde{l}'}\right)^*$ and $g_{Lij}^{\chi l\tilde{l}'}$ are given in \cite{Kingman}.
In order to give order estimation we consider only the photino contribution \cite{Suzuki}

\be
d_e=-e(\alpha/24\pi)(m_e|A_e|/M_{\tilde{\gamma}}^3)\sin (\varphi_A-\varphi_{\tilde{\gamma}})f(m_{\tilde{e}}^2/M_{\tilde{\gamma}}^2).
\label{MSSM3}
\ee
Here 
\be
f(x)\equiv \frac{12}{(x-1)^2}\left(\frac{1}{2}+\frac{3}{x-1}-\frac{2x+1}{(x-1)^2}\ln x\right).
\label{MSSM4}
\ee
with $f(1)=1$.
Also $\varphi_A$ and $\varphi_{\tilde{\gamma}}$ are the phases of $A$ term and photino (Majorana) mass, respectively. If we assume $m_{\tilde{e}}\approx M_{\tilde{\gamma}}$, then \bref{MSSM3} is expressed as 
\be
d_e\approx -1.0\times 10^{-27}\times(M_{\tilde{\gamma}}^3/1\mbox{TeV})^{-3}(|A_e|/1\mbox{TeV})\sin (\varphi_A-\varphi_{\tilde{\gamma}})~\mbox{[e cm}]
\label{MSSMvalue1}
\ee
with
\be
|A_e|\approx |\mu|\tan \beta.
\ee
\begin{figure}[h]
\begin{center}
\includegraphics[scale=0.8]{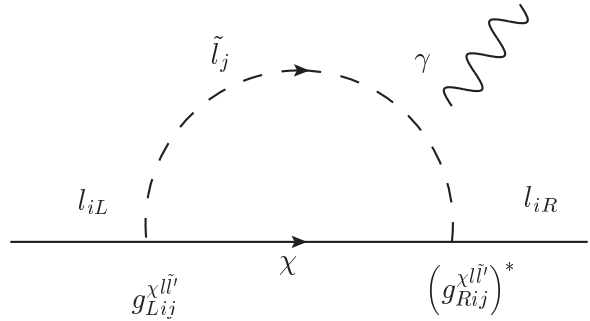}
\end{center}  
  \caption{MSSM diagrams giving rise to lepton EDM. $\chi$ indicates neutralino and chargino.
}\label{MSSM}
\end{figure}
Using \bref{d_n}, nucleon EDM is estimated similarly as \cite{Suzuki}
\bea
d_n\approx& -&e(2\alpha_s/81\pi)(m_d|A_d|/M_{\tilde{g}}^3)\sin (\varphi_{Ad}-\varphi_{\tilde{g}})f(M_{\tilde{d}}^2/M_{\tilde{g}}^2)\nonumber\\
&-&e(2\alpha_s/81\pi)(m_u|A_u|/M_{\tilde{g}}^3)\sin (\varphi_{Au}-\varphi_{\tilde{g}})f(M_{\tilde{u}}^2/M_{\tilde{g}}^2).
\label{MSSMvalue2}
\eea
Substituting $m_u=2.3$MeV, $m_d=4.8$ MeV, $\alpha_s=0.1$ and assuming the universality \\
$|A_u|=|A_d|,~\varphi_{Au}=\varphi_{Ad},~m_{\tilde{u}}=m_{\tilde{d}}=M_{\tilde{g}}$, we obtain
\be
d_n=-1\times 10^{-25}(M_{\tilde{g}}/1\mbox{TeV})^{-3}(|A_d|/1\mbox{TeV})\sin(\varphi_{Ad}-\varphi_{\tilde{g}})~\mbox{[e cm]}.
\label{MSSMvalue3}
\ee
$\clubsuit$ {\bf Condition;}\\
As is easily seen from \bref{MSSMvalue1} and \bref{MSSMvalue3}, the peaks of SUSY around  $10^{-25}$ for neutron and $10^{-27}$ [e cm] for electron, in Fig.1 correspond to those in the units of respective expressions.

Usually, SUSY predictions on EDM is described as \cite{Engel}
\be
d_f\approx e\frac{m_f}{\Lambda^2}\frac{\alpha_k}{4\pi}\sin \phi_{CPV}
\label{order1}
\ee
However, as we have represented above, EDMs have not only different qualitative predictions but also different constraints from obserbations in lepton and hadrons.
Indeed LHC gives the severe constraints on hadronic parts and the recent lower bound of strongly interaction sparticle masses is in the CMSSM with  $\tan\beta=30,~A_0=-2m_0$, sgn($\mu) >0$ \cite{ATLASsusy}\cite{CMSsusy},
\bea
\label{MSSMvalue4}
&&m_{\tilde{q}}=M_{\tilde{g}}>1.7~\mbox{TeV for}~ m_{\tilde{q}}=M_{\tilde{g}},\nonumber\\
&&M_{\tilde{g}}>1.3~\mbox{TeV for all}~m_{\tilde{q}},\\
&&m_{\tilde{q}}>1.6~\mbox{TeV for all}~M_{\tilde{g}}.\nonumber
\eea
You may compare \bref{MSSMvalue1} and \bref{MSSMvalue3} with \bref{order1}. $\Lambda$ is different in lepton and quark, and $\phi_{CPV}$ appears as the difference of those of $A$ term and gaugino.

\subsection{SUSY SO(10) Grand Unified Theories (GUTs)}
As we have discussed MSSM in the previous section, it has so many undetermined parameters, generically 105 new parameters. This deficit is recovered by SUSY GUT. Among many candidates, $SO(10)$ \cite{Fritzsch} is the smallest simple gauge group 
 under which the entire SM matter contents of each generation are unified into a single anomaly-free irreducible representation, ${\bf 16}$. 
The ${\bf 16}$-dimensional spinor representation in $SO(10)$ includes the right-handed neutrino and no other exotic matter particles.
Especially renormalizable minimal SUSY SO(10) (minimal SO(10)) GUT has only ${\bf 10}$ and $\overline{{\bf 126}}$ Higgs fields in Yukawa coupling and is very predictive \cite{Fuku2}.  It fixes all Yukawa couplings including left and right-handed Majorana neutrino mass matrices as well as quark and lepton mass matrices consistently. Only exceptional mismatch is large $\theta_{13}$ of the MNS mixing matrix which exceeded the upper limit at that time. However the afterward observed value was less than factor 2 of our prediction \cite{okada}. (We will discuss this remained mismatch shortly after.) This is in contrast with the fact that almost all the other phenomenological models predicted values at least one order less than the observed value. 
Neutrinos have masses and we must modify \bref{ss} to incorporate 
\be
\tilde{N}_R{\bf A}_D\tilde{L}H_u
\ee
with
\be
{\bf A}_D=A_0{\bf Y}_D
\ee
in \bref{USB3}. Here $Y_D$ is the Yukawa coupling of the Dirac neutrino which is tightly connected with the other quark-lepton mass matrices. $N_R$ is the heavy right-handed neutrino, which induces light left-handed neutrino via the type I seesaw mechanism. Thus, unlike the MSSM, minimal SO(10) GUT predicts all Yukawa couplings and unifies quark-leptons at GUT level and its predictions has much less ambiguities than MSSM.
EDMs of the charged-leptons are given by
\begin{eqnarray}
 d_{\ell_i}/e =  - m_{\ell_i} 
  \Im \left(A_L^{ii}-A_R^{ii} \right) \; . 
\end{eqnarray}
Here we have changed the normalization of $A_{L,R}$ from \bref{imaginary} by $\frac{em_l}{2}$. These complex $A_{L,R}$ are induced through  the renormalization effects in the same manner as for the lepton flavour violation (LFV) processes. 
\begin{figure}[h]
\begin{center}
\includegraphics[scale=0.6]{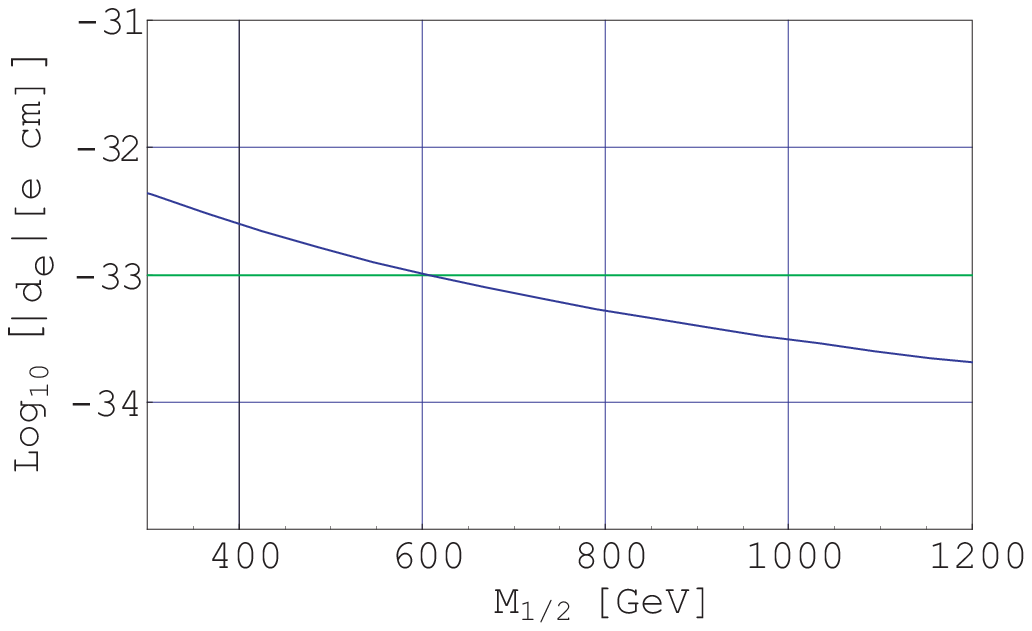}
\includegraphics[scale=0.6]{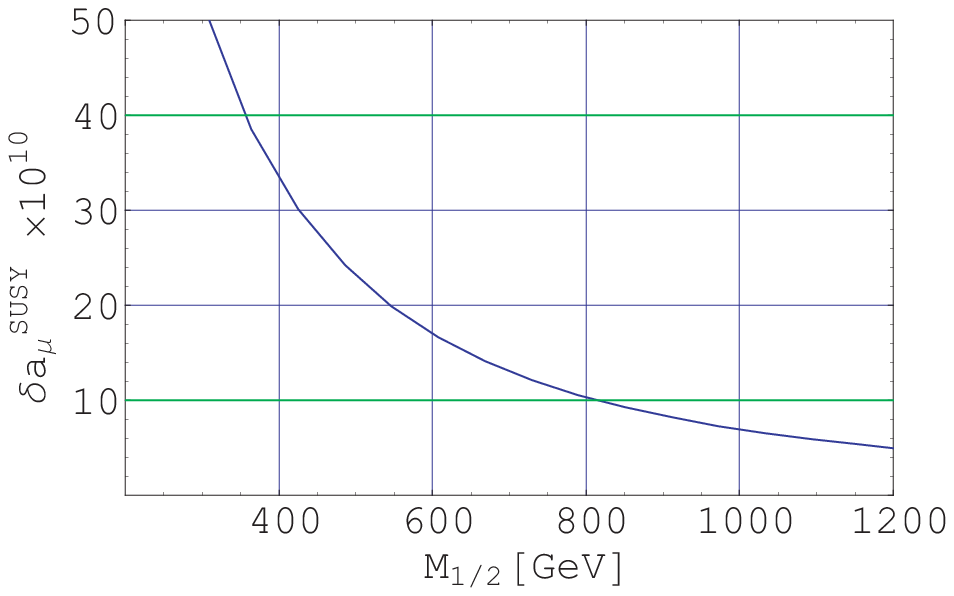}
\includegraphics[scale=0.6]{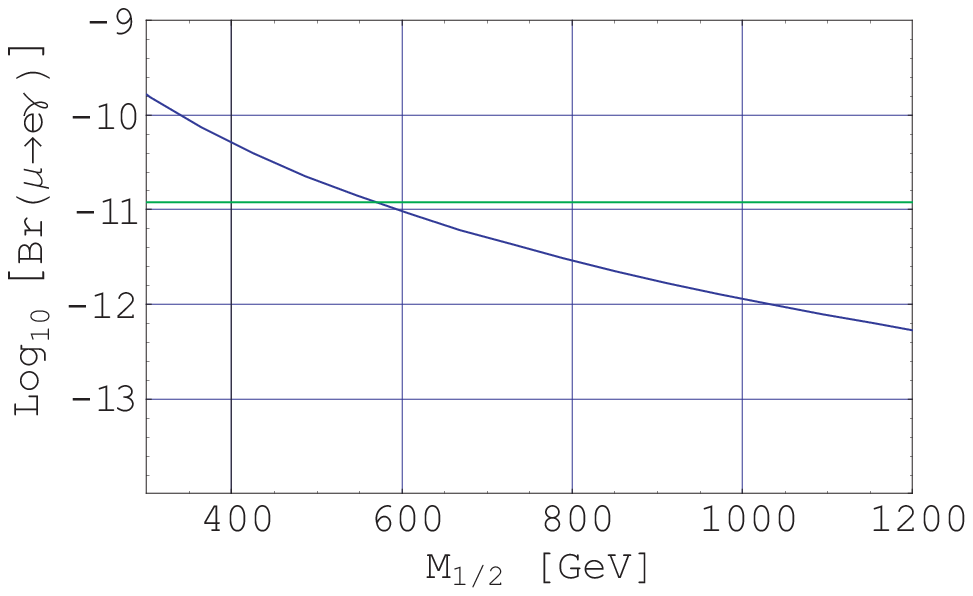}
\includegraphics[scale=0.6]{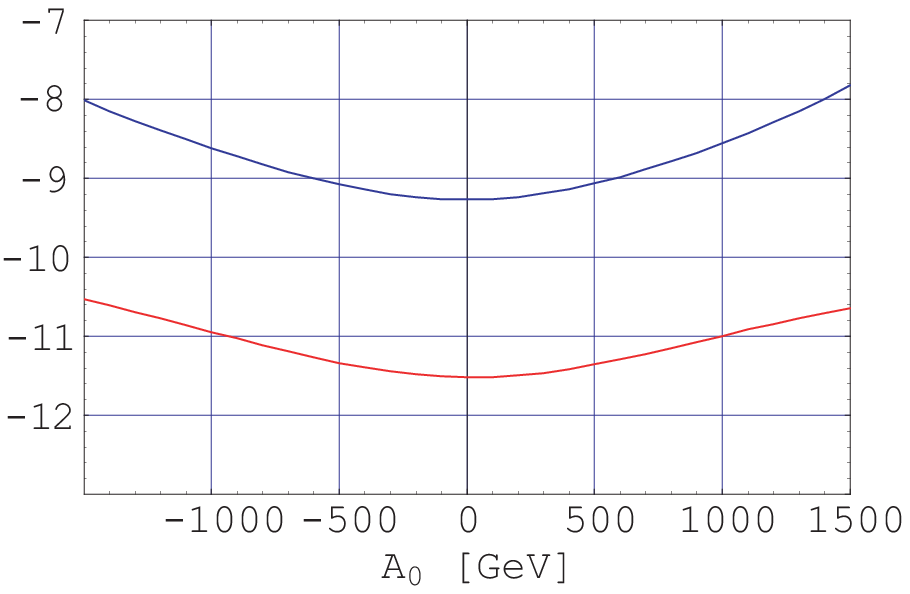}
\caption{
 The predictions for the electron EDM $|d_e|$,
the muon anomalous MDM $\delta a_\mu$,
and the decay branching ratio of $\mu \to e \gamma$
in the minimal SUSY SO(10)
with respect to the universal gaugino mass $M_{1/2}$. Trilinear term $A_0$ is assumed to be zero except for the last panel. The horizontal line just above $10^{-11}$ in the third Br($\mu\to e\gamma$) is the experimental upper limit at that time. See \bref{mutoegamma} for the present limit. These three panels are taken from \cite{Fuku3}. The last panel is added for reference to see the behaviour of non zero $A_0$, where
the branching ratios,  
$\mbox{Br}(\tau \rightarrow \mu \gamma)$ (top)
and 
$\mbox{Br}(\mu \rightarrow e \gamma)$ (bottom), are given
 as functions of $A_0$ (GeV) 
 for $m_0 =600$ GeV and $M_{1/2}=800$ GeV \cite{fukureview}.}
\label{fig:eEDM_MSSO10}
\end{center}
\end{figure}

Here the primary source of the CP-violation 
 is the CP-phases in the $Y_D$, and we obtained the results 
 of the electron and muon EDMs 
 such as (we find $d_e, d_{\mu} < 0$) (Fig. 5) \cite{Fuku3},
\be
 |d_e| = 10^{-34}-10^{-33}~ \mbox{[e cm]~ and}~|d_{\mu}| = 10^{-31} - 10^{-30} ~\mbox{[e cm]}.
\label{SO10value}
\ee
These values are still far below the present experimental upper bounds, 
 $ d_e \leq 8.7 \times 10^{-29}$ [e cm] 
 and $ d_{\mu} \leq 1.9 \times 10^{-19}$ [e cm]. 
 We presented the electron EDM in Fig. 5 as a function of $M_{1/2}$ 
 along the cosmological constraint valid at high tan$\beta$ and null $A_0$ \cite{Lahanas},
\begin{eqnarray}
m_0(\mbox{GeV}) = \frac{9}{28} M_{1/2}(\mbox{GeV}) + 150(\mbox{GeV}). 
 \label{relation} 
\end{eqnarray}
We have presented several predictions other than the EDMs since the minimal SO(10) is so predictive that many predictions are tightly connected to each others.  
We are now undertaking a revised fitting of data by taking into account the discovery of $126$ GeV Higgs particle and the negative searches for SUSY particles. We will discuss on the implications of the recent experiments in the next section.\\
$\clubsuit$ {\bf Condition;}\\
The predictions of Fig.5 used the parameters which satisfied the low energy quark-lepton mass spectra, mixing angles etc.
However, these fittings were performed before the discovery of Higgs particle. BSM physics depends on the SUSY breaking boundary conditions and SUSY breaking mechanism. We adopted CMSSM with gauge mediation ($A_0=0$), which gives $m_h=119$ GeV with $\tan\beta=45$. The correct $126$ GeV is obtained by modifying  $A_0\to -3000$ GeV from zero. In \cite{okada} we fitted all quark-lepton mass matrices including Dirac neutrino Yukawa and heavy  Majorana neutrino mass matrices with some simplified parameters (numbers of parameters are 15 for type I seesaw). That is, mass eigen values are assumed to be real and positive and seesaw mechanism type-I. If we adopt parameters fully generically (numbers of parameters are 21 for type I + II seesaw), not only mismatch of $\theta_{13}$ mentioned above is remedied but also the consistencies concerned with gauge coupling unification may be solved \cite{mimura}. 
The discrepancy of $R(D)$ and $R(D^*)$ mentioned in 2HDM can be discussed in this model.
We have the effective dimension-five interactions \cite{Meljanac},  
\be
-W_5 =  C_L^{ijkl} \, \frac{1}{2} q_i q_j q_k \ell_l
+ C_R^{ijkl} \, u^c_i e^c_j u^c_k d^c_l ,
\label{dim5}
\ee
which also induces the dangerous proton decay.  Here we obtain the explicit form of the Wilson coefficients of $C_L^{ijkl},~ C_R^{ijkl} $. The ful consistencies including proton decay are now under consideration.  As for the diphoton excess at $750$ GeV is very interesting since the mass of additional neutral Higgs scalars ($H_0, ~A_0$) may be around $750$ GeV.
However, we have not introduced exotic fermions. We need more precise measurement of this process since the background is not fully clarified yet.
As is easily seen from Fig.5, the predictions of minimal SO(10) model are very sensitive to $m_0,~M_{1/2},~A_0$, so near-future advanced experiments of EDM, LFV are expected to confine this model.

\section{The connections to the LHC and the Other Experiments}
Here we give some comments on the recent results by the CERN Large Hadronic Collider (LHC) and other groups.
On July 2012 the LHC groups announced the discovery of Higgs-like particle around 126 GeV \cite{ATLAS} \cite{CMS}.
This is not only the discovery of the last unknown particle in the SM but also gives the serious impacts to the BSM physics, especially to SUSY.
In this paper we have discussed the conditions under which new BSM models give large EDMs relative to those of the SM. 
We briefly explain why $126$ GeV Higgs mass is serious for SUSY. 
For tree level, Higgs mass satisfies an inequality
\be
m_h < M_Z|\mbox{cos}(2\beta)|
\ee
with $M_Z=91.2$GeV, which obviously disagrees with reality \footnote{This is the case for the MSSM. For the case of supersymmetric LR model, it relaxes to $m_h\leq \sqrt{2}M_W$ if the $SU(2)_R$ breaking scale is of order TeV \cite{Babu}.}. One-loop correction to $m_h$ in CMSSM is \cite{Harber2}\cite{Okada}\cite{Ellis2}
\be
m_h^2\approx M_Z^2\mbox{cos}^22\beta+\frac{3}{4\pi^2}\frac{m_t^4}{v^2}\left[\mbox{ln}\frac{M_S^2}{m_t^2}+\frac{X_t^2}{M_S^2}\left(1-\frac{X_t^2}{12M_S^2}\right)\right],
\label{higgs1}
\ee
where
\be
M_S=\sqrt{m_{\tilde{t}_1}m_{\tilde{t}_2}},~~X_t=A_t-\mu\mbox{cot}\beta,~~v=174\mbox{GeV}
\ee
with the trilinear Higgs-stop coupling constant $A_t$. So $126$ GeV points to heavy stop masses and/or large $X_t$ (left-right mixing).
Experimental null observation for SUSY particles would also give large sfermion masses in the CMSSM scenario.
One-loop EDM contribution in the MSSM is roughly proportional to $O(M_S^{-2})$ and heavy $M_S$ reduces EDM.
In Fig.~\ref{fig:eEDM_MSSO10} we have assumed $A_0=0$ in the first three panels since $A_0$ appears at two-loop correction in the gauge mediation SUSY breaking adopted in \cite{Fuku3} and is suppressed.
In order to preserve rather small $M_S$ for the hierarchy problem and still give large loop correction in \bref{higgs1} to produce $126$ GeV Higgs, we may take rather large $A_0$. However, the situation is not so simple. For instance,  since soft masses and $A$-term contribute quite differently in different processes, we must reinforce many assumptions of unknown parameters made in a model by applying it to many different phenomena.
Let us first consider the anomalous MDM of muon,
$a_\mu \equiv (g-2)/2$.
 The muon anomalous MDM
has been measured very precisely%
~\cite{Bennett} as
\begin{eqnarray}
a_\mu^{\text{exp}} = 11659208.0(6.3)\times 10^{-10},
\end{eqnarray}
where the number in parentheses shows $1\sigma$ uncertainty, which is $3.5 \sigma$ deviating from the SM \cite{HLMNT}. In SO(10) model, we can adjust this deviation as is shown in the third panel of Fig. 5 and, therefore, can not change the parameters freely.  This model also constrains the LFV processes. An approximate formula of the LFV decay rate 
 \cite{Hisano-etal} reads as
\begin{eqnarray}
\Gamma (\ell_i \rightarrow \ell_j \gamma)& =&\frac{e^2}{16\pi}m_{l_i}^5\left(|A_L^{ij}|^2+|A_R^{ij}|^2\right)\nonumber\\
&\sim&  \frac{e^2}{16 \pi} m_{\ell_i}^5 
 \times  \frac{\alpha_2}{16 \pi^2} 
 \frac{\left| \left(\Delta  m^2_{\tilde{\ell}} \right)_{ij}\right|^2}{M_S^8} 
 \tan^2 \beta \; .
 \label{LFVrough}
\end{eqnarray}
Here $ \left(\Delta  m^2_{\tilde{\ell}} \right)_{ij}$ 
 is the slepton mass estimated as
\begin{eqnarray}
 \left(\Delta  m^2_{\tilde{\ell}} \right)_{ij}
 \sim - \frac{3 m_0^2 + A_0^2}{8 \pi^2} 
 \left( Y_{\nu}^{\dagger} L Y_{\nu} \right)_{ij} \; .  
 \label{leading}
\end{eqnarray}
$Y_\nu$ is the Dirac Yukawa coupling matrix and the distinct thresholds of the right-handed 
 Majorana neutrinos are taken into account by the matrix $ L = \log [M_{\rm GUT}/M_{R_i}] \delta_{ij}$. $Y_\nu$ and $M_{R_i}$ are explicitly determined in the minimal SO(10) GUT model.

Thus a large $A_0$, as we have shown in the fourth panel in Fig. 5, exceeds the up-to-date upper bound for $\mu\to e\gamma$,
\be
Br(\mu\to e\gamma)<4.2\times 10^{-13} ~\mbox{at 90\%C.L. \cite{MEG} },
\label{mutoegamma}
\ee
in the given parameter values. This is remedied in \cite{mimura}. EDM experiments also give rather severe constraints on models.
Some extended SUSY models assert that the present upper limit for $d_e$ indicates PeV scale sfermion masses \cite{Altmannshofer}\cite{Nath}. However, these claims are due to additional assumtions of mass hierarchy or additional SUSY particles beyond the simple extension of CMSSM. As we have mentioned, muon $g-2$ requires rather small smuon mass, which is in contrast with rather large stop mass for the $126$ GeV Higgs particle. This, together with many others, gives some tensions in the simple CMSSM model. Addition of new vectorlike matter is one of extensions to relax these tensions \cite{Okada-Moroi} \cite{Gogoladze}. However, such an extension is not well confirmed and the prediction of PeV sfermion is, therefore, only one of possibilities at the present stage.

\section{Discussion}
In this letter, we have tried to fill the gaps between theory model predictions and their prerequisites in electric dipole moments, clarifying the conditions under which those predictions are given. 
Arguments have been made on the typical BSM models, namely 2HDM, LR symmetric model, MSSM, and SUSY SO(10) GUT models. \bref{demu}, \bref{LRvalue1}, \bref{LRvalue2}, \bref{MSSMvalue1}, \bref{MSSMvalue3}, and \bref{SO10value} are their results. Almost all the BSM models have many parameters which can not be wholly determined only by the compatibility with experiments and by the self-consistency of their models. Their numerical results, therefore, are given after setting some further assumptions on these undetermined parameters. In this sence, we must be careful on the conditions or the assumptions made behind the various predictions. Understanding models correctly is indispensable to the improvements of model building. At the same time, these assumptions must be continuously reinforced by applying them to many different kinds of phenomena. 
\section*{Acknowledgements}
The works of T.F. and K.A are supported in part by Grant-in-Aid for Scientific Research (A) from Japan Society for the Promotion of Science (No.26247036).

\end{document}